\documentclass[default]{sn-jnl}

\jyear{2021}%

\newcommand{\Fn}{\mathbb F_{q^n}}
\newcommand{\Fq}{\mathbb F_{q}}
\newcommand{\fq}{\mathbb F_{q}}
\newcommand{\q}{\mathrm{q}}

\newcommand{\F}{\mathbb F}
\newcommand{\qt}{\llbracket}
\newcommand{\tq}{\rrbracket}

\usepackage{enumitem}
\setlist[enumerate]{
	labelsep=8pt,
	labelindent=0pt,
	itemindent=0pt,
	leftmargin=*,
	before=\setlength{\listparindent}{-\leftmargin},
}

\theoremstyle{thmstyleone}%
\newtheorem{theorem}{Theorem}
\newtheorem{proposition}[theorem]{Proposition}%

\theoremstyle{thmstyletwo}%
\newtheorem{example}{Example}%

\theoremstyle{thmstylethree}%
\usepackage{mathtools}

\begin{document}

\title[]{Encoding and Decoding of Several Optimal Rank Metric Codes}


\author*[1]{\fnm{Wrya K.} \sur{Kadir}}\email{wrya.kadir@uib.no}

\author[1]{\fnm{Chunlei} \sur{Li}}\email{Chunlei.li@uib.no}

\author[2]{\fnm{Ferdinando} \sur{Zullo}}\email{ferdinando.zullo@unicampania.it}

\affil*[1]{\orgdiv{Department of Informatics}, \orgname{University of Bergen}, \orgaddress{\street{}\city{Bergen}, \postcode{5008}, \country{Norway}}}

\affil[2]{\orgdiv{Dipartimento di Matematica e Fisica}, \orgname{University of Campania “Luigi Vanvitelli”},  \orgaddress{\street{Viale Lincoln, 5}, \city{Caserta}, \postcode{81100}, \country{Italy}}}



\abstract{This paper presents encoding and decoding algorithms for several families of optimal rank metric codes whose codes are in restricted forms of symmetric, alternating and Hermitian matrices. First, we show the evaluation encoding is the right choice for these codes and then we provide easily reversible encoding methods for each family. Later unique decoding algorithms for the codes are described. The decoding algorithms are interpolation-based and can uniquely  correct errors for each code with rank up to  $\lfloor(d-1)/2\rfloor$ in polynomial-time, where $d$ is the minimum distance of the code. }

\keywords{Rank Metric Codes, Hermitian Matrices, Symmetric Matrices, Alternating Matrices, Linearized Polynomials, Interpolation-based Decoding, Berlekamp-Massey Algorithm}



\maketitle
\section{Introduction}


Rank metric codes were introduced first by Delsarte in \cite{Delsarte:1978aa}, and independently by Gabidulin in \cite{Gabidulin} and Roth in \cite{roth1991maximum}. They have been extensively investigated because of their applications in crisscross error correction \cite{roth1991maximum}, cryptography \cite{gabidulin1991ideals} and network coding \cite{silva2008rank}.
The coding-theoretic properties of these codes have been studied in detail, and constructions of optimal codes with respect to a Singleton-like bound, known as MRD codes, have been found. An interested reader may refer to \cite{GorlaRavagnani,Sheekey2019} for more details.

Known decoding algorithms for MRD codes can be generally classified in two different approaches: syndrome-based decoding  as in \cite{Gabidulin1985,GPT,roth1991maximum,Richter} and interpolation-based decoding as in \cite{Loidreau:2006aa,Tovohery2018,Kadir-li20,Li2019,Li2018,kadir2021new}.
Gabidulin in \cite{Gabidulin1985} solves the key equation in  the decoding process by employing the linearized version of \textit{extended Euclidean (LEE) algorithm}, while in \cite{Richter}, the key equation was solved by a linearized version of \textit{Berlekamp-Massey (BM) algorithm}. The error values in both decoding algorithms in \cite{Gabidulin1985} and \cite{Richter} are computed by an algorithm called \textit{Gabidulin algorithm}. Loidreau in \cite{Loidreau:2006aa} proposed the first interpolation-based decoding approach for MRD codes and considered the analogue of \textit{Welch-Berlekamp (WB) algorithm}, which was  originally used to decode Reed-Solomon codes \cite{welch_Ber_1986error}. The algorithm directly  gives the code's interpolation polynomial and computing the error vector is not required in the decoding process.

In \cite{Sheekey}, Sheekey proposed the first family of MRD codes over $\Fn$ which is linear over $\Fq$ (instead of $\Fn$ as the well-known Gabidulin codes) and his idea were used later to introduce new MRD codes that are linear over a sub-field of $\Fn$ \cite{lunardon2018,Otal2017,Otal:2018aa,Otal2018,TrombettiZhou2019}.  When the rank of the error vector reaches the maximum unique decoding radius, syndrome-based decoding approach works only for MRD codes that are linear over the main extension field. 
Randrianarisoa in \cite{rosenthal2017decoding,Tovohery2018} , gave an interpolation based decoding algorithm for twisted Gabidulin codes. Later this idea was adopted to decode additive generalized twisted Gabidulin codes and Trombetti-Zhou rank metric codes \cite{Kadir-li20,Kadir-li-Zullo21}. Again BM algorithm is involved in the process of solving the key equations in \cite{Kadir-li20} and \cite{Kadir-li-Zullo21} and it reduces the decoding problem to the problem of solving  the projective polynomial equation  $x^{q^v+1}+ax+b=0$ and quadratic polynomial equation $x^2+cx+d=0$ over $\Fn$, respectively. A similar idea is also used in \cite{kadir2021new} to decode Gabidulin codes beyond half the minimum distance. All the decoding algorithms described above have polynomial-time complexities. The result in \cite{silva2009fast} shows that when low-complexity normal basis are used, the complexity can be reduced even further. Solving the key equations carried out by BM or LEE algorithm are the most expensive steps in the above decoding algorithms. 

Besides the aforementioned new MRD codes, there are also some restricted rank metric codes that are linear over a subfield of $\Fn$ which are not defined based on Sheekeys' idea. 
The study of subsets of \emph{restricted} matrices equipped with rank metric was started in 1975 by Delsarte and Goethals in \cite{delsarte1975alternating}, in which they considered sets of alternating bilinear forms. The theory developed in \cite{Delsarte:1978aa} and \cite{delsarte1975alternating} found applications also in the classical coding theory. Indeed, the evaluations of the forms found in \cite{delsarte1975alternating} give rise to subcodes of the second-order Reed-Muller codes, including the Kerdock code and the chain of Delsarte–Goethals codes; see also \cite{schmidt2010symmetric}.

Using the theory of association schemes, bounds, constructions and structural properties of restricted rank metric codes have been investigated in symmetric matrices \cite{longobardi2020automorphism,schmidt2015symmetric,zhou2020equivalence},  alternating matrices \cite{delsarte1975alternating} and    Hermitian matrices \cite{schmidt2018hermitian,trombetti2020maximum}.

In this paper we will present both encoding and decoding algorithms for  several optimal  symmetric,  alternating and  Hermitian rank metric codes. Since the targeted codes are not linear over the extension field, syndrome-based decoding algorithms in \cite{Gabidulin1985} is not applicable. We choose interpolation-based decoding approach which is able to decode errors up to half of the minimum distance in polynomial time for all the aforementioned codes. 
A part of our work in this paper responds to a suggestion in \cite{de2021hermitian}, where the authors suggested studying the decoding of Hermitian rank metric codes.

\section{Preliminaries}\label{sec:pre}

Let $\mathbb{F}_{\q}$ denote a finite field of $\q$ elements and 
$\mathbb{F}_{\q}^{n\times n}$ be the set of the square matrices of order $n$ defined over $\mathbb{F}_{\q}$.
We can equip $\mathbb{F}_{\q}^{n\times n}$ with the following metric
\[ d_r(A,B)=\mathrm{rk} (A-B), \]
where $\mathrm{rk}(A-B)$ is the rank of the difference matrix $A-B$.
If $\mathcal{C}$ is a subset of $\mathbb{F}_{\q}^{n\times n}$ with the property that 
\[d=\min\{\mathrm{rk} (A-B) \colon A,B \in \mathcal{C}, A \ne B\},\] 
then $\mathcal{C}$ is called a \emph{rank metric code with minimum distance} $d$, or that $\mathcal{C}$ is a $d$-\emph{code}, see e.g.\ \cite{schmidt2010symmetric}.
A rank metric code $\mathcal{C}$ is said to be \textit{additive} if it is closed under the classical matrix addition $+$ and said to be \textit{linear} over a subfield $\mathbb{E}$ of $\mathbb{F}_{\q}$ if it is closed under both matrix addition and scalar multiplication by any element in $\mathbb{E}$.

Let $\mathcal{L}_{n, \q}$ denote the quotient $\F$-algebra of all \textit{$\q$-polynomials} over $\F_{\q^n}$ with degree smaller than $n$ , namely,
\[\mathcal{L}_{n,\q}=\left\{ \sum\limits_{i=0}^{n-1} a_i x^{\q^i} \colon a_i \in \F_{\q^n} \right\}. \]
It is well known that the $\mathbb{F}_{\q}$-algebra $\mathcal{L}_{n, \q}$ is actually isomorphic to the $\mathbb{F}_{\q}$-algebra $\mathbb{F}_\q^{n\times n}$.
Hence many rank metric codes $\mathcal{C} \subseteq \mathbb{F}_\q^{n\times n}$ are expressed in terms of $\q$-polynomials in $\mathcal{L}_{n, \q}$. If $\q$ is fixed or the context is clear, we can use the term \textit{linearized polynomials} instead of $\q$-polynomials. 

Here we recall one important property of the Dickson matrix associated with $\q$-polynomials which is critical for the decoding in this paper.

\begin{proposition}\label{prop-Dickson-tovo}
	Let  $L(x)=\sum\limits_{i=0}^{n-1}a_ix^{\q^i}$ over $\F_{\q^n}$ be a $\q$-polynomial with rank $t$. Then its associated Dickson matrix 
	\small
	\begin{equation}\label{EqDicksonmatrix}
	D=\begin{pmatrix}
	a_{i-j({\,\rm mod }n)}^{\q^i}
	\end{pmatrix}_{n\times n}=\begin{pmatrix*}[l]
	a_0& a_{n-1}^{\q}& \cdots& a_1^{\q^{n-1}}\\
	a_1& a_0^{\q}&\cdots& a_2^{\q^{n-1}}\\
	\vdots&\vdots&\ddots&\vdots\\
	a_{n-1}& a_{n-2}^{\q}&\cdots& a_{0}^{\q^{n-1}}
	\end{pmatrix*},
	\end{equation}
	\normalsize
	has rank $t$ over $\F_{\q^n}$ and any $t\times t$ submatrix formed by $t$ consecutive rows and $t$ consecutive columns in $D$ is non-singular. 
\end{proposition}
For the first part of the above result see \cite{MENICHETTI-86,dickson-book}, whereas for the last part we refer to \cite{Tovohery2018}.

Below we shall introduce three families of rank metric codes whose codewords have restrictive forms. The first two consist of symmetric and alternating matrices over $\F_q$, respectively, and the third one consists of Hermitian matrices defined over $\F_{q^2}$, where $q$ is a prime power.

\smallskip


Recall that a matrix $A\in \mathbb{F}_{q}^{n\times n} $ is said to be \textit{symmetric} if $A^T=A$ and is said to be \textit{alternating} if $A^T=-A$, where $A^T$ is the transpose matrix of $A$.
Let ${\rm S}_{n}(q)$ and ${\rm A}_{n}(q)$ be the set of all symmetric matrices and alternating matrices of order $n$ over $\mathbb{F}_{q}$, respectively. 
Following the connection given in \cite{longobardi2020automorphism}, 
the set ${\rm S}_n(q)$ can be identified as 
\[{\mathcal{S}}_n(q)=\left\{ \sum\limits_{i=0}^{n-1} c_i x^{q^{i}} \colon c_{n-i}=c_i^{q^{n-i}} \,\text{ for } \, i \in \{0,\ldots,n-1\} \right\}\subseteq \mathcal{L}_{n,q}. \]
The set ${\rm A}_n(q)$ can be identified as 
\[{\mathcal{A}}_n(q)=\left\{ \sum\limits_{i=0}^{n-1} c_i x^{q^{i}} \colon c_{n-i}=-c_i^{q^{n-i}} \,\text{ for } \, i \in \{0,\ldots,n-1\} \right\}\subseteq \mathcal{L}_{n,q}. \]

Consider the conjugation map $\overline{\cdot} $ from $\mathbb{F}_{q^2}$ to itself: $x \mapsto \overline{x} = x^q$. For a matrix $A\in \mathbb{F}_{q^2}^{n\times n}$, we denote by $A^*$ the conjugate transpose of $A$, which is obtained by applying the conjugate map to all entries of $A^{T}$.
Recall that a matrix $A\in \mathbb{F}_{q^2}^{n\times n}$ is said to be \emph{Hermitian} if $A=A^*$.
Let  ${\rm H}_{n}(q^2)$  be the set of all Hermitian matrices of order $n$ over $\mathbb{F}_{q^2}$.
Similarly, it can be identified as 
\[{\mathcal{H}}_n(q^2)=\left\{ \sum\limits_{i=0}^{n-1} c_i x^{q^{2i}} \colon c_{n-i+1}=c_i^{q^{2n-2i+1}} \,\text{ for } \, i \in \{0,\ldots,n-1\} \right\}\subseteq \mathcal{L}_{n,q^2},\]
where the indices are taken modulo $n$. Note that if $n$ is odd then $c_{(n+1)/2}$ belongs to $\mathbb{F}_{q^n}$.

It is can be easily verified that that these three sets, together with the classical sum of matrices and the scalar multiplication by elements in $\Fq$, are $\Fq$-vector spaces  with dimensions
\[ \dim_{\fq}({\rm S}_{n}(q))=\frac{n(n+1)}{2},\,\,\,\dim_{\fq}({\rm A}_{n}(q))=\frac{n(n-1)}{2},\,\,\,\dim_{\fq}({\rm H}_{n}(q^2))=n^2. \]
A subset of ${\rm S}_{n}(q)$, ${\rm A}_{n}(q)$ or ${\rm H}_{n}(q^2)$ endowed with the rank distance will be termed a symmetric, alternating or Hermitian rank metric code, respectively, or symmetric, alternating or Hermitian $d$-code if $d$ is the minimum distance of the considered code. 
With the isomorphism between $ \mathbb{F}_\q^{n\times n}$ and $\mathcal{L}_{n, \q}$, $\q\in \{q, q^2\}$, the codewords in these restricted rank metric codes will be represented in polynomials throughout this paper.
For simplicity, we will denote by $x^{[i]}:=x^{q^i}$ and $x^{\qt i\tq} := x^{q^{2i}}$ for any integer $i$.

\subsection{Optimal Symmetric and Alternating $d$-Codes}\label{sec:sym-alt}

For symmetric and  alternating rank metric codes, the following bounds on their size have been established \cite{delsarte1975alternating,schmidt2015symmetric}.


\begin{theorem}\label{th:boundsym}\cite[Theorem 3.3]{schmidt2015symmetric}
	Let $\mathcal{C}$ be a symmetric $d$-code in $\F_q^{n\times n}$. 
	If $d$ is even, suppose also that $\mathcal{C}$ is additive. Then
	\[ \#\mathcal{C} \leq \left\{ \begin{array}{ll}
	q^{n(n-d+2)/2} & \text{if}\,\,\, n-d\,\,\text{is even},\\
	q^{(n+1)(n-d+2)/2} & \text{if}\,\,\, n-d\,\,\text{is odd}.
	\end{array}\right. \]
\end{theorem}


\begin{theorem}\label{th:boundalter}\cite[Theorem 4]{delsarte1975alternating}
	Let $m=\lfloor \frac{n}2 \rfloor$.
	Let $\mathcal{C}$ be an alternating $2e$-code in $\F_q^{n\times n}$. 
	Then
	\[ \#\mathcal{C} \leq q^{\frac{n(n-1)}{2m}(m-e+1)}. \]
\end{theorem}
A symmetric (resp. alternating) $d$-code is said to be  \textit{optimal} if its parameters satisfy the equality in Theorem \ref{th:boundsym} (resp.  Theorem \ref{th:boundalter}).
The following theorems present some instances of optimal symmetric (resp. alternating) $d$-codes, where $\mathrm{Tr }_{q^n/q}(x)=x+x^q+\cdots+x^{q^{n-1}}$ is the trace function from $\F_{q^n}$ to $\Fq$.
%

\begin{theorem}\cite[Theorem 4.4]{schmidt2015symmetric}\label{Th-Schmidt-4.4}
	Let $n$ and $d$ be two positive integers such that $1\leq d\leq n$ and $n-d$ is even. 
	The symmetric forms 
	$S:\F_{q^{n}}\times \F_{q^{n}}\rightarrow \F_{q}$ given by $S(x,y)=\mathrm{Tr }_{q^n/q}\left(yL(x)\right)$ with
	\begin{equation}\label{ex:E1s}
	L(x)= b_0x + \sum\limits_{j=1}^{\frac{n-d}{2}} \left( b_jx^{q^j}+(b_jx)^{q^{n-j}}  \right),
	\end{equation}
	as $b_0,\ldots,b_{\frac{n-d}{2}}$ range over $\F_{q^{n}}$,
	form an additive optimal $d$-code in $\mathrm{S}_n(q)$. 
\end{theorem}
In \cite[Theorem 4.1]{schmidt2015symmetric} it has been shown that constructions of optimal symmetric $d$-codes with $n-d$ odd in $\mathrm{S}_n(d)$ can be obtained by puncturing the examples of optimal symmetric $d$-codes found in \cite[Theorem 4.4]{schmidt2015symmetric}.


\begin{theorem}\cite[Theorem 7]{delsarte1975alternating}\label{Th-Altern}
	Let $n$ and $e$ be two positive integers such that $n$ is odd and $1\leq 2e\leq n-1$, and let $d=2e$. 
	The alternating forms 
	$A:\F_{q^{n}}\times \F_{q^{n}}\rightarrow \F_{q}$ given by $A(x,y)=\mathrm{Tr }_{q^n/q}\left(yL(x)\right)$ with
	\begin{equation}\label{ex:E1s-2}
	L(x)= \sum\limits_{j=e}^{\frac{n-1}{2}} \left( b_jx^{q^j}-(b_jx)^{q^{n-j}}  \right),
	\end{equation}
	as $b_e,\ldots,b_{\frac{n-1}{2}}$ range over $\F_{q^{n}}$,
	form an additive optimal $d$-code in $\mathrm{A}_n(q)$. 
\end{theorem}

\subsection{Optimal Hermitian $d$-Codes}\label{sec:her}

Schmidt characterized the upper bound on the size of Hermitian $d$-codes as follows \cite[Theorem 1]{schmidt2018hermitian}.

\begin{theorem}\label{th:boundHerm}\cite[Theorem 1]{schmidt2018hermitian}
	An additive Hermitian $d$-code  $\mathcal{C}$ in $\F_{q^2}^{n\times n}$ satisfies
	\[ \#\mathcal{C} \leq q^{n(n-d+1)}. \]
	Moreover, when $d$ is odd, this upper bound holds also for non-additive Hermitian $d$-codes.
\end{theorem}

A Hermitian $d$-code is called a \textit{optimal} Hermitian $d$-code if it attains the above bound. 
Schmidt in \cite{schmidt2018hermitian} also gave constructions for optimal Hermitian d-codes for all possible value of $n$ and $d$, except if $n$ and
$d$ are both even and $3<d<n$. There are some examples of optimal Hermitian $d$-codes, see \cite{schmidt2018hermitian,trombetti2020maximum}.
We recall two examples given in \cite[Theorems 4 and 5]{schmidt2018hermitian}, where $\mathrm{Tr }_{q^{2n}/q^2}$ is the trace function from $\F_{q^{2n}}$ to $\F_{q^2}$.
\begin{theorem}\cite[Theorem 4]{schmidt2018hermitian}\label{Th-Schmidt-4}
	Let $n$ and $d$ be integers of opposite parity satisfying $1\leq d\leq n$. 
	The Hermitian forms 
	$H:\F_{q^{2n}}\times \F_{q^{2n}}\rightarrow \F_{q^2}$ given by $H(x,y)=\mathrm{Tr }_{q^{2n}/q^2}\left(y^qL(x)\right)$ with
	\begin{equation}\label{ex:E1}
	L(x)= \sum\limits_{j=1}^{\frac{n-d+1}{2}} \left( (b_jx)^{q^{(2n-2j+2)}}+b_j^{q} x^{q^{(2j)}} \right),
	\end{equation}
	as $b_1,\ldots,b_{\frac{n-d+1}{2}}$ range over $\F_{q^{2n}}$,
	form an additive optimal $d$-code in $\mathrm{H}_n(q^2)$.
\end{theorem}

\begin{theorem}\cite[Theorem 5]{schmidt2018hermitian}\label{Th-Schmidt-5}
	Let $n$ and $d$ be odd integers satisfying $1\leq d\leq n$. 
	The Hermitian forms 
	$H:\F_{q^{2n}}\times \F_{q^{2n}}\rightarrow \F_{q^2}$ given by $H(x,y)=\mathrm{Tr }_{q^{2n}/q^2}\left(y^qL(x)\right)$ with
	\begin{equation}\label{ex:E2}
	L(x)= (b_0x)^{q^{(n+1)}}+ \sum\limits_{j=1}^{\frac{n-d}{2}} \left( (b_jx)^{q^{(n+2j+1)}}+b_j^{q} x^{q^{(n-2j+1)}} \right),
	\end{equation}
	as $b_0$ ranges over $\F_{q^n}$ and $b_1,\ldots,b_{\frac{n-d}{2}}$ range over $\F_{q^{2n}}$,
	form an additive optimal $d$-code in $\mathrm{H}_n(q^2)$.
\end{theorem}

\section{Encoding}\label{sec:encoding}

In the literature, no encoding  method has been given for symmetric, alternating and Hermitian $d$-codes. 	This section is dedicated to the encoding of these three types of restricted $d$-codes. 
As a matter of fact, the encoding of a optimal  $d$-code $\mathcal{C}$ is mainly concerned with setting up a one-to-one correspondence between a message space of size $\#\mathcal{C}$
and the code $\mathcal{C}$ in an efficient way, which ideally also allows for an efficient decoding algorithm.

\subsection{Encoding of symmetric $d$-codes}\label{sec:enc-symm}
We start with the encoding of the optimal symmetric $d$-codes of size $q^{n(n-d+2)/2}$  in Theorems \ref{Th-Schmidt-4.4}, where $n-d$ is even. 
The family of codes is linear over $\mathbb{F}_q$ and the message space is naturally a vector space over  $\mathbb{F}$ with dimension $n(n-d+2)/2$. But we can represent each message in the form of a $k$-dimensional vector over $\mathbb{F}_{q^n}$ where $k=(n-d+2/2)$ and the set of all the message vectors are closed under $\mathbb{F}_q$-linear operations.  

In order to present a polynomial-time decoding algorithm for the optimal symmetric $d$-codes in Theorem \ref{Th-Schmidt-4.4}, we shall express their codewords as evaluations of certain polynomials at linearly independent points over $\F_{q}$. For this reason, we need to employ a pair of dual base in $\F_{q^{n}}$ over $\F_{q}$. 
Recall that given an ordered $\F_{q}$-basis  $(\alpha_1,\ldots,\alpha_n)$ of $\F_{q^{n}}$, 
its dual basis is defined as the ordered $\F_{q}$-basis  $ (\beta_1,\ldots,\beta_n)$ of $\F_{q^{n}}$ such that 
$$
\mbox{Tr}_{q^{n}/q}(\alpha_i\beta_j) = \delta_{ij} \text{ for } i = 1, 2, \dots, n,
$$ where  $\delta_{ij}$ denotes the Kronecker delta function.
Note that a dual basis always exists for a given order basis $(\alpha_1,\ldots,\alpha_n)$ of $\F_{q^n}$  \cite[Definition 2.30]{lidl1997}.

Let $(\alpha_1,\ldots,\alpha_n)$, $(\beta_1,\ldots,\beta_n)$ be a pair of dual base of $\F_{q^{n}}$ over $\Fq$.
We will write $\mbox{Tr}_{q^n/q}(x)$ as ${\rm Tr}(x)$ for simplicity when the context is clear.
Let $L(x)$ be a linearized polynomial as in Theorem \ref{Th-Schmidt-4.4}. For the symmetric form  we have 
\[ 
S(x,y)=\mbox{Tr}(x L(y)).
\]

Now, we denote the associated matrix of $S$ with respect to the ordered $\F_{q}$-basis $(\alpha_1,\ldots,\alpha_n)$ by $\mathcal{S}$, of which the $(i,j)$-th entry   $\mathcal{S}(i,j)$ is given by 
$$\mathcal{S}(i,j)=S(\alpha_i, \alpha_j)=\mbox{Tr}(\alpha_j L(\alpha_i)) .$$

Furthermore, the codewords of the additive $d$-code in Theorem \ref{Th-Schmidt-4.4} can be expressed in the symmetric matrix form as follows: let $x, y \in \F_{q^n}$, then $x=\sum\limits_{i=1}^n x_i\alpha_i$ and $y=\sum\limits_{j=1}^n y_j\alpha_j$ for some $x_i,y_j \in \F_q$ and
\begin{align*}
S(x,y)&=\text{Tr}\left(\Big(\sum\limits_{j}y_j\alpha_j\Big)\sum\limits_{i}x_iL(\alpha_i)\right)
=\mbox{Tr}\left( \sum\limits_{i,j}x_iy_j\alpha_jL(\alpha_i)\right)\\
&=\sum\limits_{i,j}x_iy_j\mbox{Tr}\left(\alpha_jL(\alpha_i)\right)
= \sum\limits_{i,j}x_i\mathcal{S}(i,j)y_j
=(x_1,\ldots,x_n)\cdot \mathcal{S}\cdot \left(\begin{array}{c}
y_1  \\
y_2\\
\vdots\\
y_n
\end{array}\right),
\end{align*} where $\mathcal{S}(i,j)$ is the $(i,j)$-th entry in $\mathcal{S}$.

In the following we show that the evaluation of the corresponding linearized polynomial at linearly independent elements $\alpha_1,\ldots,\alpha_n$ is a proper encoding method. 

Define an $n$-dimensional vector over $\F_{q}$ as 
$$s=(s_1,\ldots,s_n)=(\beta_1,\ldots,\beta_n)\cdot \mathcal{S}^T.$$ 
Since the $i$-th row of $\mathcal{S}$ is given by $(\mathrm{Tr}(\alpha_1L(\alpha_i)), \ldots, \mathrm{Tr}(\alpha_nL(\alpha_i)))$ and since each $L(\alpha_i)$ can be written as $\sum_t c_t \beta_t$ for some $c_t \in \F_q$, we can write $s_i$ as 
\begin{align*}
s_i
&=\sum\limits_{j}\beta_j S(i,j) =\sum\limits_{j}\beta_j \mbox{Tr}(\alpha_jL(\alpha_i))
\\   & =\sum\limits_{j}\beta_j\mbox{Tr}\big(\alpha_j\sum\limits_{t}c_t\beta_t\big)
=\sum\limits_{j}\beta_j\sum\limits_{t}c_t\mbox{Tr}(\alpha_j\beta_t)
\\&     =\sum\limits_{j}\beta_jc_j=L(\alpha_i)
\end{align*}
since ${\rm Tr}(x)$ is linear over $\F_{q}$ and $(\beta_1,\ldots,\beta_n)$ is the dual basis of $(\alpha_1,\ldots, \alpha_n)$.
From the equality $s_i=L(\alpha_i)$, we see that the encoding of symmetric $d$-codes given by ${\rm Tr}(yL(x))$, as in Theorems \ref{Th-Schmidt-4.4} and \ref{Th-Altern}, can be seen as the evaluation of 
$L(x)$ at the basis $\alpha_1, \ldots, \alpha_n$ of $\F_{q^{n}}$. 

\smallskip
With the above preparation, we are now ready to look at the encoding of the optimal symmetric $d$-codes in Theorem \ref{Th-Schmidt-4.4} more explicitly. 

Let $\omega_0, \ldots, \omega_{n-1}$ be a  basis of  $\F_{q^{n}}$ over $\mathbb{F}_q$.  
For optimal symmetric $d$-codes in Theorem \ref{Th-Schmidt-4.4}, the linearized polynomial can be expressed as  $$L(x) = b_0x+\sum\limits_{j=1}^{k-1} \left(b_jx^{[j]}+(b_jx)^{[n-j]} \right),$$ 
where $k=(n-d+2)/2$.  
Then the encoding of
a message $f=(f_0,\ldots, f_{k-1}) \in \F_{q^n}^k$ for the symmetric codes in Theorem  \ref{Th-Schmidt-4.4} can be expressed as the evaluation of the following linearized polynomial at points $\omega_0, \ldots, \omega_{n-1}$:
\begin{equation*}
\begin{array}{rcl}
L(x) 
&=& 
f_0x+\left( \sum\limits_{j=1}^{k-1} f_j x^{[j]} + 
(f_jx)^{{[n-j]}}\right) = \sum\limits_{i=0}^{n-1}\tilde{f}_ix^{[i]},
\end{array}
\end{equation*}
where 
\begin{equation}\label{eq-tilde-f-symmetric}
\begin{array}{rcl}
\tilde{f}&=(\tilde{f}_0, \dots,\tilde{f}_{k-1},0,\ldots,0 ,\tilde{f}_{n-k+1},\ldots,\tilde{f}_{n-1}) \\&= (f_0, \dots, f_{k-1}, 0, \dots, 0, f^{[n-k+1]}_{k-1}, \dots, f^{[n-1]}_{1}).
\end{array}
\end{equation}

Let $	N=
\begin{pmatrix}
\omega_i^{[j]}
\end{pmatrix}_{n\times n}$
be the $n\times n$ \textit{Moore matrix} generated by $\omega_i$'s. 
So the encoding of optimal symmetric and optimal alternating $d$-codes  can be expressed as \begin{equation}\label{eq1-1}
(f_0,\ldots,f_{k-1})\mapsto (L(\omega_0),\ldots,L(\omega_{n-1}))=\tilde{f}\cdot N^T,
\end{equation}
where $\tilde{f}=(\tilde{f}_0, \ldots, \tilde{f}_{n-1})$ and $N^T$ is the transpose of the matrix $N$.  
Note that the first $k$ and the last  $k-1$ elements of $\tilde{f}$ are nonzero. This means at most $n-d+1$ columns of the matrix $N^T$ are involved in the encoding process. 

\subsection{Encoding of alternating $d$-codes}\label{sec:enc-alt}
The encoding of alternating $d$-codes in Theorem \ref{Th-Altern} can be done similarly since the codewords in $A(x,y)$ has the same form ${\rm Tr}(yL(x))$ as in Theorem \ref{Th-Schmidt-4.4}.

For alternating  $d$-codes in Theorem \ref{Th-Altern}, the linearized polynomial can be expressed as 
$$L(x) = \sum\limits_{j=e}^{\frac{n-1}{2}} \left(b_jx^{[j]}-(b_jx)^{[n-j]} \right).$$ 
Note that in Theorem \ref{Th-Altern}, the parameters $n$ is odd and $d=2e$. The optimal alternating codes are $\mathbb{F}_q$-linear with dimension $n(n-d+1)/2$. For simplicity, we again consider the message vectors in the form of vectors over $\mathbb{F}_{q^n}$.

Let $\omega_0, \ldots, \omega_{n-1}$ be a basis of  $\F_{q^{n}}$ over $\mathbb{F}_q$. The encoding of
	a message $f=(f_0,\ldots, f_{k-1}) \in \F_{q^n}^k$ can be expressed as the evaluation of the following linearized polynomial at points $\omega_0, \ldots, \omega_{n-1}$:
	\begin{equation*}
	\begin{array}{rcl}
	    L(x) 
	    &=& 
	     \left( \sum\limits_{j=e}^{\frac{n-1}{2}} f_{j-e} x^{[j]} - 
	    (f_{j-e}x)^{{[n-j]}}\right)=\sum\limits_{i=0}^{n-1}\tilde{f}_{i}x^{[i]},
  \end{array}
	\end{equation*} where 
	\begin{equation}\label{eq-tilde-f-alter}
\begin{array}{rcl}
\tilde{f}&=(0, \dots,0,\tilde{f}_{e},\ldots,\tilde{f}_{\frac{n-1}{2}},\tilde{f}_{\frac{n+1}{2}},\ldots ,\tilde{f}_{n-e},0,\ldots,0) \\&= (0,\dots,0, f_{0},  \dots, f_{k-1}, -f^{[\frac{n+1}{2}]}_{k-1}, \dots, -f^{[n-e]}_{0},0,\ldots,0).
\end{array}
	\end{equation}

 Similarly, the encoding of optimal alternating $d$-code  can be expressed as \begin{equation}\label{eq1-1-1}
   (f_0,\ldots,f_{k-1})\mapsto (L(\omega_0),\ldots,L(\omega_{n-1}))=\tilde{f}\cdot N^T,
\end{equation}
where $\tilde{f}=(\tilde{f}_0, \ldots, \tilde{f}_{n-1})$ and $N^T$ is the transpose of the matrix $N$.  As shown in \eqref{eq-tilde-f-alter},  at most $n-d+1$ columns of the matrix $N$ are involved in computation.

\subsection{Encoding of Hermitian $d$-codes}\label{sec:enc-her}

	This section is dedicated to the encoding of the optimal Hermitian $d$-codes of size $q^{n(n-d+1)}$ explained in Theorems \ref{Th-Schmidt-4} and \ref{Th-Schmidt-5}. 
	Given positive integers $d, n$ with $1\leq d\leq n$, for encoding of optimal Hermitian $d$-codes we are going to  set up a one-to-one correspondence between a message space of size $q^{n(n-d+1)}$, and Hermitian optimal $d$-codes, which later permits us  to decode efficiently. Therefore, for a message space of size $q^{n(n-d+1)}$, we may assume its elements as
	 vectors over $\F_{q^n}$ of dimension $k=n-d+1$. 
	
	For the optimal Hermitian $d$-codes in Theorems \ref{Th-Schmidt-4} and \ref{Th-Schmidt-5}, we shall express their codewords as evaluations of certain polynomials at linearly independent points over $\F_{q^2}$. For this reason, we need to introduce the Hermitian variant of a basis in $\F_{q^{2n}}$ over $\F_{q^{2}}$. 
	Given an ordered $\F_{q^2}$-basis  $(\alpha_1,\ldots,\alpha_n)$ of $\F_{q^{2n}}$, 
	its \textit{Hermitian dual basis} is defined as the ordered $\F_{q^2}$-basis  $( \beta_1,\ldots,\beta_n)$ of $\F_{q^{2n}}$ such that 
	$$
	\mbox{Tr}_{q^{2n}/q^2}(\alpha_i^q\beta_j) = \delta_{ij} \text{ for } i = 1, 2, \dots, n,
	$$ where $\mbox{Tr}_{q^{2n}/q^2}$ is the relative trace function from $\F_{q^{2n}}$ to $\F_{q^2}$, namely, $\mbox{Tr}_{q^{2n}/q^2}(x)=\sum\limits_{i=0}^{n-1}x^{q^{2i}}$ and $\delta_{ij}$ denotes the Kronecker delta function.
	Note that such a Hermitian dual basis always exists for a given order basis $(\alpha_1,\ldots,\alpha_n)$. Indeed, since there exist a dual basis $(\gamma_1,\ldots,\gamma_n)$ 
	for $(\alpha_1,\ldots,\alpha_n)$ satisfying $\mbox{Tr}_{q^{2n}/q^2}(\alpha_i\gamma_j)=\delta_{ij}$, one can simply takes $\beta_j = \gamma_j^{q^{2n-1}}$ for $j=1,2,\dots, n$ and then the above Hermitian dual property follows.
	We shall also write $\mbox{Tr}_{q^{2n}/q^2}()$ as $\mbox{Tr}()$ for simplicity whenever there is no ambiguity.

	Let $(\alpha_1,\ldots,\alpha_n)$ be an $\F_{q^2}$-basis of $\F_{q^{2n}}$  and $( \beta_1,\ldots,\beta_n)$ be its Hermitian dual as described above.
	Let $x,y \in \F_{q^{2n}}$, then $x=\sum\limits_{i=1}^n x_i \alpha_i$ and $y=\sum\limits_{i=1}^n y_i \beta_i$, for some $x_i,y_i \in \F_{q^2}$. It is clear that  $\mathrm{Tr}(x^qy)=\sum\limits_{i,j=1}^nx_i^qy_j\mathrm{Tr}(\alpha_i^q\beta_j)=\sum\limits_{i=1}^nx_i^qy_i=\langle(x_1^q,\ldots,x_n^q),(y_1,\ldots,y_n)\rangle$. 
	
	\smallskip
	
	Note that the Hermitian forms in Theorems \ref{Th-Schmidt-4} and \ref{Th-Schmidt-5}  are of the form
	$
	H(x,y)=\mbox{Tr}(x^q L(y))
	$.
	Now, we denote the associated matrix of $H$ with respect to the ordered $\F_{q^2}$-basis $(\alpha_1,\ldots,\alpha_n)$ by $\mathcal{H}$, of which the $(i,j)$-th entry   $\mathcal{H}(i,j)$ is given by 
	$$\mathcal{H}(i,j)=H(\alpha_i, \alpha_j)=\mbox{Tr}(\alpha_j^q L(\alpha_i)) .$$

	Furthermore, the codewords of the additive $d$-code in Theorem \ref{Th-Schmidt-5} can be expressed in the Hermitian matrix form as follows
	\begin{align*}
	H(x,y)&=\text{Tr}\left(\Big(\sum\limits_{j}y_j\alpha_j\Big)^q\sum\limits_{i}x_iL(\alpha_i)\right)
	=\mbox{Tr}\left( \sum\limits_{i,j}x_iy_j^q\alpha_j^qL(\alpha_i)\right)\\
	&=\sum\limits_{i,j}x_iy_j^q\mbox{Tr}\left(\alpha_j^qL(\alpha_i)\right)
	= \sum\limits_{i,j}x_i\mathcal{H}(i,j)y_j^q
	=(x_1,\ldots,x_n)\cdot \mathcal{H}\cdot \left(\begin{array}{c}
	y_1^q  \\
	y_2^q\\
	\vdots\\
	y_n^q
	\end{array}\right),
	\end{align*} where $\mathcal{H}(i,j)$ is an element in $\mathcal{H}$.
 In the following we show that the evaluation of the corresponding linearized polynomial at linearly independent elements $\alpha_1,\ldots,\alpha_n$ is a proper encoding method. 
	Define an $n$-dimensional vector over $\F_{q^{2}}$ as 
	$$h=(h_1,\ldots,h_n)=(\beta_1,\dots,\beta_n)\cdot \mathcal{H}^T.$$ 
	Since the $i$-th row of $\mathcal{H}$ is given by $(\mathrm{Tr}(\alpha_1^qL(\alpha_i)), \ldots, \mathrm{Tr}(\alpha_n^qL(\alpha_i)))$ and since each $L(\alpha_i)$ can be written as $\sum_t c_t\beta_t$ for some $c_t \in \F_{q^2}$, we can write $h_i$ as 
	\begin{align*}
	h_i
	&=\sum\limits_{j}\beta_j H(i,j) =\sum\limits_{j}\beta_j \mbox{Tr}(\alpha_j^qL(\alpha_i))
	\\   & =\sum\limits_{j}\beta_j\mbox{Tr}\big(\alpha_j^q\sum\limits_{t}c_t\beta_t\big)
	=\sum\limits_{j}\beta_j\sum\limits_{t}c_t\mbox{Tr}(\alpha_j^q\beta_t)
	\\& =   \sum\limits_{t}\beta_tc_t=L(\alpha_i),
	\end{align*}
	where and the fourth and fifth equality signs hold because
	${\rm Tr}(x)$ is linear over $\F_{q^2}$ and $(\beta_1,\ldots,\beta_n)$ is the Hermitian dual basis of $(\alpha_1,\ldots, \alpha_n)$. 
	From the equality $h_i=L(\alpha_i)$, we see that the encoding of Hermitian $d$-codes given by ${\rm Tr}(y^qL(x))$, as in Theorems \ref{Th-Schmidt-4} and \ref{Th-Schmidt-5}, can be seen as the evaluation of 
	$L(x)$ at the basis $\alpha_1, \ldots, \alpha_n$ of $\F_{q^{2n}}$. 
	
	\smallskip
	
	With the above preparation, we are now ready to look at the encoding of the Hermitian $d$-codes in Theorems \ref{Th-Schmidt-4} and \ref{Th-Schmidt-5} more explicitly.

Let $\kappa=\lceil\frac{n-d}{2}\rceil$ and $H$ be the Hermitian form given in Theorem \ref{Th-Schmidt-4}. 
The linearized polynomial in \eqref{ex:E1}
can be written as 
$$L(x) = \sum\limits_{j=1}^{\kappa} \left( (b_jx)^{\qt n+1-j\tq}+b_j^{q} x^{\qt j\tq } \right),$$
and assuming $m=\frac{n+1}{2}$, similarly one can write the linearized polynomial in \eqref{ex:E2} as

$$L(x) = (b_0x)^{\qt m\tq } + \sum\limits_{j=1}^{\kappa} \left( (b_jx)^{\qt m+j\tq }+b_j^{q} x^{\qt m-j\tq} \right).$$ 

Let $\{1,\eta\}$ be an $\F_{q^n}$-basis of $\F_{q^{2n}}$.
Let $\alpha_0,\alpha_1, \ldots, \alpha_{n-1}$ be a basis of  $\F_{q^{2n}}$ over $\mathbb{F}_{q^2}$.
Raising all the basis elements $\alpha_i$ to the $q^2$-th power will still give a linearly independent set of elements in $\mathbb{F}_{q^{2n}}$. We use $\alpha_0^{q^2},\alpha_1^{q^2}, \ldots, \alpha_{n-1}^{q^2}$ as the evaluation points for optimal Hermitian $d$-codes in Theorem \ref{Th-Schmidt-4}. The reason for this is to keep the consistent form $L(x)=l_0x^{\qt 0 \tq}+l_1x^{\qt 1 \tq}+\cdots+l_{n-1}x^{\qt n-1 \tq}$ for the linearized polynomial representation (Employing $\alpha_0,\ldots,\alpha_{n-1}$ as the evaluation points for this codes will obligate us to use the linearized polynomial of the form  $L(x)=l_0x^{\qt 1 \tq}+l_1x^{\qt 2 \tq}+\cdots+l_{n-1}x^{\qt n \tq}$).

The encoding of
a message $f=(f_0,\ldots, f_{k-1}) \in \F_{q^n}^k$ can be expressed as the evaluation of the following linearized polynomial at points $\alpha_0^{q^2},\alpha_1^{q^2}, \ldots, \alpha_{n-1}^{q^2}$:

\begin{equation}\label{eq-tilde-f-Her1}
\begin{array}{rcl}
L(x) 
&=& 
\left( \sum\limits_{j=0}^{\kappa-1} (f_j+\eta f_{\kappa+j})^qx^{\qt n-1-j \tq} + 
(f_j+\eta f_{\kappa+j}x)^{\qt j\tq }\right)
=\sum\limits_{i=0}^{n-1}\tilde{f}_ix^{\qt i \tq},
\end{array}
\end{equation}

where 
\begin{equation}\label{eq-tilde-f-her1}
\begin{array}{rcl}
\tilde{f}&=(\tilde{f}_0, \dots,\tilde{f}_{\kappa-1},0,\ldots,0,\tilde{f}_{n-\kappa},\ldots, \tilde{f}_{n-1}) 
= ((f_0+\eta  f_{\kappa})^{\qt 0\tq},\dots,\\&(f_{\kappa-1}+ \eta f_{2\kappa-1})^{\qt \kappa-1\tq},0,  \dots, 0,
(f_{\kappa-1}+ \eta f_{2\kappa-1})^{q},\ldots,(f_0+ \eta f_{\kappa})^{q}),    
\end{array}
\end{equation}
and $k=2\kappa$.
For the optimal Hermitian $d$-code in Theorem \ref{Th-Schmidt-5} and the evaluation points $\alpha_0,\alpha_1, \ldots, \alpha_{n-1}$, the encoding of
a message $f=(f_0,\ldots, f_{k-1}) \in \F_{q^{n}}^k$ can be expressed as the evaluation of the following linearized polynomial at points $\alpha_0,\alpha_1, \ldots, \alpha_{n-1}$:
\begin{equation}\label{eq-tilde-f-Her2}
\begin{array}{rcl}
L(x)&= (f_0x)^{\qt m \tq} + \left( \sum\limits_{j=1}^{\kappa} (f_j+\eta f_{\kappa+j})^qx^{\qt m-j\tq} +((f_j+ \eta f_{\kappa+j})x)^{\qt m+j\tq}\right)
\\&=\sum\limits_{i=0}^{n-1}\tilde{f}_ix^{\qt i\tq},
\end{array}
\end{equation}
where 
\begin{equation}\label{eq-tilde-f-her2}
\begin{array}{rcl}
\tilde{f}=&(0, \dots,0,\tilde{f}_{m-\kappa},\ldots,\tilde{f}_{m-1},\tilde{f}_{m},\tilde{f}_{m+1},\ldots,\tilde{f}_{m+\kappa},0,\ldots, 0)
\\=& (0,\dots,0,(f_{\kappa}+\eta f_{2\kappa})^q,\ldots,(f_{1}+\eta f_{\kappa+1})^q,f_0^{\qt m \tq},\\&(f_1+\eta f_{\kappa+1})^{\qt m+1\tq},\ldots,(f_{\kappa}+\eta f_{2\kappa})^{\qt m+\kappa\tq},0,\ldots,0 ), 
\end{array}
\end{equation}
and $k=2\kappa+1$.

Let $	M_l=
\begin{pmatrix}
\alpha_i^{\qt j+l\tq }
\end{pmatrix}_{n\times n}$
be the $n\times n$ \textit{Moore matrix} generated by $\alpha_0^{q^{2l}},\alpha_1^{q^{2l}}, \ldots, \alpha_{n-1}^{q^{2l}}$
where $l\in \{0,1\}$. We take $l=1$ when we consider $\alpha_0^{q^{2}},\alpha_1^{q^{2}}, \ldots, \alpha_{n-1}^{q^{2}}$   as the evaluation points which is used in \eqref{eq-tilde-f-Her1} and $l=0$ when $\alpha_0,\alpha_1,\ldots,\alpha_{n-1}$ are the evaluation points in \eqref{eq-tilde-f-Her2}.

So the encoding of the optimal Hermitian rank metric codes can be expressed as \begin{equation}\label{eq1}
(f_0,\ldots,f_{k-1})\mapsto (L(\alpha_0^{q^{2l}}),\ldots,L(\alpha_{n-1}^{q^{2l}}))=\tilde{f}\cdot M_l^T,
\end{equation}
where $\tilde{f}=(\tilde{f}_0,\ldots,\tilde{f}_{n-1})$  and $M_l^T$ is the transpose of the matrix $M_l$.

When $n,d$ are integers with opposite parities as shown in \eqref{eq-tilde-f-Her1}, only the first $\kappa$ and the last $\kappa$ elements of $\tilde{f}$ are non zero. Also in the case when $n,d$ are both odd integers, as can be seen in \eqref{eq-tilde-f-Her2}, the first $m-\kappa$  and the last $ m-\kappa-2$ elements of $\tilde{f}$ are zero. So we only use $n-d+1$ columns of the Moore matrix in the encoding process.

In summary, the encoding of the optimal symmetric, alternating and Hermitian $d$-codes relies on converting the codewords of those codes to simplified linearized polynomials $L(x)$ under carefully-chosen base of the extension fields, which enables us to treat encoding of those codes as evaluations of $L(x)$ at linearly independent points.

\section{Decoding}\label{sec:decoding}
 
In Section 3 the encodings of symmetric, alternating and Hermitian $d$-codes are in the form of polynomial evaluation. In this section we will present interpolation-based decoding of those codes, which make use of some nice properties of Dickson matrices
in Proposition \ref{prop-Dickson-tovo}.
 
\subsection{Key equations for error interpolation polynomials} \label{sec:dec-keyeq}

We start with the optimal symmetric and alternating $d$-codes in Theorems \ref{Th-Schmidt-4.4} and \ref{Th-Altern}. Note that their codewords are in the form ${\rm Tr}(yL(x))$ and can be deemed as $n$-dimensional
vectors $(L(\omega_0), \dots, L(\omega_{n-1}))$ over $\F_{q^n}$. We assume errors that occur in transmission or storage medium are also vectors in $\F_{q^n}^n$.

Given a message $f=(f_0,\dots, f_{k-1})\in\F_{q^n}^k$, its corresponding codeword $c=\tilde{f}\cdot N^T$, where $\tilde{f}$ and $N^T$ are as given in Section 3.
Let $r=(r_0, \dots, r_{n-1})$ over $\F_{q^n}$ be a received word when the codeword $c\in \F_{q^n}^n$ is transmitted, namely, $r=c+e$ for certain error vector $e \in \F_{q^n}^n$.
Suppose $g(x)=\sum\limits_{i=0}^{n-1}g_ix^{[i]}$ is the error interpolation polynomial such that
	\begin{equation}\label{EqInterpolation-2}
	g(\omega_i)=e_i=r_i-c_i, \quad i=0, \ldots, n-1.
	\end{equation}
	Clearly the error vector  $e$ is uniquely determined by  the error interpolation polynomial $g(x)$, and vice versa. Denote $\tilde{g}=(g_0,\ldots, g_{n-1})$. Then it follows that
	\begin{equation}\label{r=c+e-1}
	    	r = c+e = (\tilde{f}+\tilde{g}) N^T.
	\end{equation}

Denote by $G$ the associated Dickson matrix of the $q$-polynomial $g(x)$, i.e., 
\begin{equation*}\label{Eq-DM-G-Simplified-1}
G=\begin{pmatrix}g^{[ j] }_{i-j~({\rm mod~}n)} 
\end{pmatrix}_{n\times n}
= \left(G_0 \,\, G_1 \,\, \ldots \,\, G_{n-1}\right)
=\begin{bmatrix}
g_0 & g_{n-1}^{[1]}& \ldots & g_1^{[n-1]}\\
g_1 & g_{0}^{[1]} & \ldots & g_2^{[n-1]}\\
\vdots & \vdots & \ddots & \vdots \\
g_ {n-1} & g_{n-2}^{[1]} & \ldots & g_{0}^{[n-1]}
\end{bmatrix},
\end{equation*} 
where the subscripts are taken modulo $n$. 
Suppose the error $e$ has rank $t$, by Proposition \ref{prop-Dickson-tovo} we know that $G$ has rank $t$ and any $t\times t$ submatrix formed by $t$ consecutive rows and columns in $G$ has rank $t$.
Furthermore, the first column of $G$ can be expressed as a linear combination of $G_1, \dots, G_t$ as 
\begin{equation}\label{Eq-G0}
G_0 = \lambda_1G_1+\cdots + \lambda_{t}G_t,
\end{equation} where $G_1,\dots, G_t$ are linearly independent over $\F_{q^n}$.

In the following we will make use of the pattern of $L(x)$ in Theorems \ref{Th-Schmidt-4.4} and \ref{Th-Altern}, which have consecutive $d-1$ zero coefficients (up to a cyclic shift on the coefficients), and the properties of $G$ in recovering the vector $\tilde{g}$.

\subsubsection{Optimal symmetric $d$-codes in Theorem \ref{Th-Schmidt-4.4}}\label{sec:dec-symm}
    
    For optimal symmetric $d$-codes, by \eqref{eq-tilde-f-symmetric} we can rewrite \eqref{r=c+e-1}  as
	\begin{align*}
	  r \cdot (N^T)^{-1}=&(\tilde{f}_0,\ldots,\tilde{f}_{k-1},0, \ldots,0,\tilde{f}_{n-k+1},\ldots,\tilde{f}_{n-1}) 
	  \\+& (g_0, \ldots,g_{k-1}, g_{k},\ldots, g_{n-k},g_{n-k+1}, \ldots, g_{n-1}).
		\end{align*}
		where $\tilde{f}_0=f_0^{[0]}$, 
		$\tilde{f}_{j} = f_j$ and $\tilde{f}_{j}=\tilde{f}_{n-j}^{[n-j]}$ for $j=1, \ldots, k-1$.  
	Recall that $k=(n-d+2)/2$ for symmetric $d$-codes in Theorem \ref{Th-Schmidt-4.4}. Letting $\beta=(\beta_0, \ldots, \beta_{n-1})=r \cdot (N^T)^{-1}$,  we obtain 
\begin{equation}\label{Eq-interpolation1-3}
g_i =
 \begin{cases}
 	     \beta_i &\text{ for } i= k, \dots, k+d-2,\\
 	     \beta_{i} - \tilde{f}_{i} & \text{ for } i=n-k+1, \dots, n-2k+1,
 \end{cases}
\end{equation} 
where the subscripts are taken modulo $n$.
%
Since the elements $g_{k},\ldots, g_{n-k}$  are known, from \eqref{Eq-G0} we can have the following system of linear equations:
\begin{equation}\label{key-eq-sys}
g_{i} = \lambda_1 g^{[1]}_{i-1} + \lambda_2 g^{[2]}_{i-2} + \cdots + \lambda_t g^{[t]}_{i-t}, \,\, k+t \leq i\leq n-k,
\end{equation} 
which contains $t$ unknowns $\lambda_1, \ldots, \lambda_{t}$ in $d-1-t$ linear equations. 

\subsubsection{Optimal alternating $d$-codes in Theorem \ref{Th-Altern}} \label{sec:dec-alt}

From \eqref{eq-tilde-f-alter} it follows that \eqref{r=c+e-1} is equivalent to 
\begin{align*}
	  r \cdot (N^T)^{-1}=&(0,\ldots,0,\tilde{f}_e,\ldots,\tilde{f}_{n-e},0,\ldots,0 )
	  \\+& (g_0, \ldots,g_{e-1}, g_{e},\ldots, g_{n-e},g_{n-e+1}, \ldots, g_{n-1}).
		\end{align*}
		where  $\tilde{f}_{j+e}=f_j$ and $\tilde{f}_{n-e+j}=-f_j^{[n-e-j]}$ for $j=0,\ldots,k$. Suppose we have $\beta=(\beta_0, \ldots, \beta_{n-1})=r \cdot (N^T)^{-1}$, similarly we obtain
		\begin{equation}\label{known-alter}
		g_i =
		\begin{cases}
		\beta_i &\text{ for } i =n-e+1, \dots, n+e-1\\
		\beta_{i} - \tilde{f}_{i} & \text{ for } i=e,\ldots, n-e,
		\end{cases}
		\end{equation}
		where the subscripts are taken modulo $n$.
	Based on \eqref{known-alter}, we obtain the following linear system of equations 
	\begin{equation}\label{key-eq-alt}
g_{i} = \lambda_1 g^{[1]}_{i-1} + \lambda_2 g^{[2]}_{i-2} + \cdots + \lambda_t g^{[t]}_{i-t}, \,\, n-e+1+t \leq i<n+e \mod n
\end{equation} 
with $t$ unknowns $\lambda_1.\ldots,\lambda_t$ in $2e-1-t=d-1-t$ linear equations.

\smallskip

From the above analysis, one sees that the equations \eqref{key-eq-sys} and \eqref{key-eq-alt} are the key equations for decoding optimal symmetric and optimal alternating $d$-codes, respectively.

\subsubsection{Optimal Hermitian $d$-codes}\label{sec:dec-her}

The approach of establishing the key equations in decoding Hermitian $d$-codes is similar to that for symmetric and alternating $d$-codes. Because Hermitian $d$-codes are defined over $\mathbb{F}_{q^2}$ instead of  $\mathbb{F}_{q}$,
we briefly describe the process in the sequel.

Suppose a Hermitian codeword $c \in \F_{q^{2n}}^n$ is transmitted and 
	a word $r=c+e$, with an error $e$ with rank $t$ added to the codeword $c$, is received. 
	Suppose $g(x)=\sum\limits_{i=0}^{n-1}g_ix^{\qt i\tq}$ is an error interpolation polynomial with rank $t$ such that
	\begin{equation}\label{EqInterpolation}
	g(\alpha_i^{\qt l \tq})=e_i=r_i-c_i, \quad i=0, \ldots, n-1 \mbox{ and }l\in \{0,1\},
	\end{equation}
	where we use $l=1$ for the Hermitian $d$-codes in Theorem \ref{Th-Schmidt-4} and $l=0$ for the codes in Theorem \ref{Th-Schmidt-5}. 
	It is clear that the error vector  $e=(e_0,\ldots,e_{n-1})$ is uniquely determined by the polynomial $g(x)$.
	Denote by $$
	G=(G_0, \dots, G_{n-1}) = 
	\begin{pmatrix}
	g^{\qt j \tq }_{i-j~({\rm mod~}n)} 
	\end{pmatrix},
	$$
	the Dickson matrix associated with $g(x)$, then $G$ has rank $t$ and we can express 
	\begin{equation}\label{Eq-G0-H}
	G_0 = \lambda_1G_1+\cdots + \lambda_{t}G_t,
	\end{equation}
	with unknown $\lambda_i$'s in $\F_{q^{2n}}$.
			
			Denote $\tilde{g}=(g_0,\ldots, g_{n-1})$. From
	\eqref{eq1} and \eqref{EqInterpolation} it follows that
	\begin{equation}\label{r=c+e}
	    	r = c+e = (\tilde{f}+\tilde{g}) M_l^T.
	\end{equation}

\noindent\textbf{Case 1.} This case considers the optimal Hermitian $d$-codes in Theorem \ref{Th-Schmidt-4}.
Recall that in Theorem \ref{Th-Schmidt-4}  the Hermitian $d$-codes have parameters $n,d$ with opposite parities and the message space was represented in $k$-dimensional vectors over $\mathbb{F}_{q^n}$ which are closed under $\mathbb{F}_{q}$-linear operations. 
Denoting $\kappa = \lceil \frac{n-d}{2}\rceil$, we can rewritten  \eqref{r=c+e} as 
	\begin{align*}
	  r \cdot (M_1^T)^{-1}=&(\tilde{f}_0,\ldots,\tilde{f}_{\kappa-1},0, \ldots,0,\tilde{f}_{n-\kappa},\ldots,\tilde{f}_{n-1}) 
	  \\& + (g_0, \ldots,g_{\kappa-1}, g_{\kappa},\ldots, g_{n-\kappa-1},g_{n-\kappa}, \ldots, g_{n-1}).
		\end{align*}
		where for $j=0,1, \ldots, \kappa-1$, 
		$\tilde{f}_{n-j-1} = (f_j+\eta f_{\kappa+j})^q  \text{ and } 
\tilde{f}_{j} = \tilde{f}_{n-j-1}^{q^{2j+1}}$, and $\{1, \eta\}$ is an $\F_{q^n}$-basis of $\F_{q^{2n}}$.

	Let $\beta=(\beta_0, \ldots, \beta_{n-1})=r \cdot (M_1^T)^{-1}$. Since $2\kappa = n-d+1$, 
we have $n-\kappa -1 = \kappa + d-2$ and
\begin{equation}\label{Eq-interpolation2-2}
g_i =
\begin{cases}
\beta_i &\text{ for } i =\kappa, \dots, \kappa+d-2\\
\beta_{i} - \tilde{f}_{i} & \text{ for } i=n-\kappa, \dots, n+\kappa-1,
\end{cases}
\end{equation}
This together with \eqref{Eq-G0-H}  gives a system of $d-1-t$ linear equations over $\mathbb{F}_{q^{2n}}$ with $t$ unknowns $\lambda_i$'s in $\F_{q^{2n}}$.

\noindent\textbf{Case 2.} This case considers the optimal Hermitian $d$-codes in Theorem \ref{Th-Schmidt-5}.
In this case $n,d$ are both odd integers. Denote $m=(n+1)/2$ and $\kappa = (n-d)/2$. Note that \eqref{r=c+e} is equivalent to
	\begin{align*}
	  r \cdot (M_0^T)^{-1}=&(0,\ldots,0,\tilde{f}_{m-\kappa}, \ldots,\tilde{f}_{m+\kappa}, 0,\ldots,0) 
	  \\& + (g_0, \ldots,g_{m-\kappa-1}, g_{m-\kappa},\ldots, g_{m+\kappa},g_{m+\kappa+1}, \ldots, g_{n-1}).
		\end{align*}
		where $\tilde{f}_m = f_0^{\qt m\tq}$ and for $j=1,2, \cdots, \kappa$, 
		$\tilde{f}_{m-j} = (f_j+\eta f_{\kappa+j})^q  \text{ and } \tilde{f}_{m+j} = \tilde{f}_{m-j}^{q^{n+2j}}$.
	Denote $\beta=(\beta_0, \ldots, \beta_{n-1})=r \cdot (M_0^T)^{-1}$. Since $\kappa = (n-d)/2$, we have 
	$n-1-(m+\kappa+1)+1+(m-\kappa) = n-2\kappa-1 = d-1$ known $g_i$'s and we can
	 obtain 
\begin{equation}\label{Eq-interpolation2-1}
	     g_i =
	     \begin{cases}
	     \beta_i &\text{ for } i =m+\kappa+1, \dots, m+\kappa+d-1\\
	     \beta_{i} - \tilde{f}_{i} & \text{ for } i=m-\kappa, \dots, m+\kappa,
	     \end{cases}
\end{equation} where the subscripts are taken modulo $n$.
Similarly, this together with \eqref{Eq-G0-H}  gives a system of $d-1-t$ linear equations over $\F_{q^{2n}}$ with $t$ unknowns $\lambda_i$'s in $\F_{q^{2n}}$.

\subsection{Reconstruction of the error polynomial}\label{sec:dec-rec-error}

Recall that the error polynomials $g(x)$ for symmetric and alternating $d$-codes are $q$-polynomials over $\F_{q^{n}}$ and the one for Hermitian $d$-codes are $q^2$-polynomials over $\F_{q^{2n}}$.
Despite the difference in representation, the approach used for recover the coefficients will be the same for those error polynomials. This observation allows us to present the common procedure of reconstructing $g(x)$'s
in a unified manner.

Let $\q \in \{q, q^2\}$. Given an error polynomial $g(x)=\sum\limits_{i=0}^{n-1} \in \F_{\q^n}[x]$ with rank $t$, its associate Dickson matrix given by
$$
G=(G_0, G_1,\dots, G_{n-1}) = \begin{pmatrix}
g_{i-j({\,\rm mod }n)}^{\q^i}
\end{pmatrix}_{n\times n}
$$ also has rank $t$ and 
$
G_0 = \lambda_1G_1 + \cdots + \lambda_t G_t
$ for $t$ unknown $\lambda_i$'s in $\F_{\q^n}$, which gives rise to a linearized recurrence as 
\begin{equation}\label{Eq-LFSR}
g_L = \lambda_1 g_{L-1}^{\q} +  \lambda_2 g_{L-2}^{\q^2} + \cdots + \lambda_t g_{L-t}^{\q^t} \text{ for } L=0, 1, \dots,n-1
\end{equation} where the subscripts of $g_i$'s are taken modulo $n$.
For the optimal symmetric, alternating and Hermitian $d$-codes in Section 2, Section 4.1 has established a system of $d-1-t$ linear equations over $\F_{\q^n}$ in $t$ unknowns $\lambda_i \in \F_{\q^n}$ for each of them.

\smallskip

	According to the pattern in $G$,  we have the following major steps for recovering the coefficients $g_i$'s:
	\begin{enumerate}[label=\indent\textbf{Step \arabic*.}]
		\item derive the unknowns $\lambda_1, \ldots, \lambda_{t}$ from the $d-1-t$ linear equations given in Section 4.1 for each optimal $d$-code;
		\item use $\lambda_1, \ldots, \lambda_{t}$ to recursively compute unknown $g_i$'s in $G$.
	\end{enumerate}

\smallskip

%

		Step 1 is the critical step in the decoding process. In Step 1 one has a system of $d-1-t$ linear equations for each optimal $d$-codes with $t$ unknowns. There are two options for solving the unknowns. The first option is simply applying Gaussian elimination algorithm on the equations; and 
	the second option is to apply the modified Berlekamp-Massey algorithm in \cite{Sidorenk}. As a matter of fact,
	with the linearized recurrence in \eqref{Eq-LFSR}, the task of Step 1 becomes finding the coefficients of modified version of a linear shift register as in \cite{Sidorenk} for given $d-1$ consecutive inputs $g_i$'s for each optimal $d$-codes.
	
	For Step 2, with the recursive relation in \eqref{Eq-LFSR} can calculate the remaining unknown coefficients $g_i$'s in a sequential order.

\subsection{Reconstruction of the original message}\label{sec:dec-rec-message}

Recall that for each optimal $d$-code, it is assumed that a codeword $c$ is transmitted and a word $r=c+e$ is received. With the error polynomials $g(x)$ obtained in Section 4.2, we are directly able to derive the 
codeword $c = r-e$. With the codeword $c$, we can obtain the coefficient vector $\tilde{f}$ of the interpolation polynomial $f(x)=\sum\limits_{i=0}^{n-1}\tilde{f}_ix^{q^{ui}}$ where $u\in \{1,2\}$. One can compute $\tilde{f}=(\tilde{f}_0,\ldots,\tilde{f}_{n-1})=c\cdot (\mathcal{A}^T)^{-1}$ where $\mathcal{A}$ is the Moore matrix associated with the linearly independent evaluation points.    When the $\tilde{f}$ is obtained, we can further reconstruct the original message $f=(f_0,\ldots,f_{k-1})$ according to the encoding for each optimal $d$-code as follows: 
\begin{itemize}
	\item\textbf{Symmetric $d$-codes.}
	\[
	f=(f_0,\ldots,f_{k-1})=(\tilde{f}_0,\ldots,\tilde{f}_{k-1}).
	\]
	\item\textbf{Alternating $d$-codes.}
		\[
	f=(f_0,\ldots,f_{k-1})=(\tilde{f}_e,\ldots,\tilde{f}_{\frac{n-1}{2}}).
	\]
	\item\textbf{Hermitian $d$-codes.}
	\begin{itemize}
	    \item Case 1. When $n,d$ have different parities: for $j\in\{ 0,\ldots,\kappa-1\}$ where $k=2\kappa$  we have the following  equations 
	    
	    \begin{equation}
	        \begin{cases}
	           \tilde{f}_j&=(f_j+\eta f_{\kappa+j})^{q^{2j}}   \\
	         \tilde{f}_{n-j-1}&=(f_j+\eta f_{\kappa+j})^q .   
	        \end{cases}
 	    \end{equation}
 	    The unknown coefficients $f_j,f_{k+j}\in \mathbb{F}_{q^n}$ for $j\in \{0,\ldots,\kappa-1\}$ can be seen as the unique coordinate vector of $\tilde{f}_j^{q^{-2j}}$ (or $\tilde{f}_{n-j-1}^{q^{-1}}$)  expressed with respect to the basis $\{1,\eta\}$ of $\mathbb{F}_{q^{2n}}$ over $\mathbb{F}_{q^n}$ and can be computed directly. 
	   
	    
	    \item Case 2. When $n,d$ are both odd:
	    for $j\in 1,\ldots,\kappa-1$ where $k=2\kappa+1$  we have the following linear system of equations 
	    
	    \begin{equation}
	        \begin{cases}
	        \tilde{f}_m&=f_0^{q^{2m}}\\
	           \tilde{f}_{m+j}&=(f_j+\eta f_{\kappa+j})^{q^{2(m+j)}}   \\
	         \tilde{f}_{m-j}&=(f_j+\eta f_{\kappa+j})^q .  
	        \end{cases}
	    \end{equation}
	    
	   The coefficient $f_0$ can be computed from the first equation as $f_0=\tilde{f}_m^{q^{-2m}}$. Similar to the Case 1., the unknown coefficients $f_j,f_{j+\kappa}$ can be seen as coordinate vector of $\tilde{f}_{m-j}^{q^{-1}}$ (or $\tilde{f}_{m+j}^{q^{-2(m+j)}}$) written with respect to the basis $\{1,\eta\}$. So we can compute all the unknown coefficients $f_0,\ldots, f_{k-1}\in \mathbb{F}_{q^n}$ and recover the message.
	 
	\end{itemize}
\end{itemize}

\subsection{Summary}\label{sec:summary}
The decoding algorithms in Section \ref{sec:decoding} share some similarities and one can summarize the decoding algorithms for all the restricted codes as follows:
\begin{itemize}
\item \textbf{Input:}  a received word $r=(r_0,\ldots,r_{n-1})$ with errors of $t\leq \frac{d-1}{2}$ rank and linearly independent points $\theta_0,\ldots,\theta_{n-1}$ in $\mathbb{F}_{q^{un}}$ where $u\in \{1,2\}$.

\item \textbf{Idea:}  Reconstructing the code's interpolation polynomial $f(x)=\sum\limits_{i=0}^{n-1}\tilde{f}_ix^{q^{ui}}$  via the error interpolation polynomial $g(x)=\sum\limits_{i=0}^{n-1}g_ix^{q^{ui}}$ where  $f(\theta_i)+g(\theta_i)=c_i+e_i=r_i$.

\item \textbf{Output:} The codeword $c=r-e$.
\end{itemize}
\begin{itemize}
    \item[(1)] Compute the coefficients $\beta_i$ of the polynomial $\beta(x)=\sum\limits_{i=0}^{n-1}\beta_ix^{q^{ui}}$ where $r_i=\beta(\theta_i)$. This is equivalent to $r\cdot(\mathcal{M}^T)^{-1}$, where $\mathcal{M}$ is the Moore matrix associated with $\theta_i$'s.
    \item[(2)] Specify the known coefficients  $(g_j,\ldots, g_{j+d-2})=(\beta_j,\ldots, \beta_{j+d-2})$, where the subscripts are taken modulo $n$, based on the code. 
    \item[(3)] Use the $2t$ known coefficients $g_i$ as the initial state in the BM algorithm and find the unique connection vector $\lambda=(\lambda_1,\ldots,\lambda_t)$.
    \item[(4)] Let $G$ be the Dickson matrix associated with $g(x)$ with rank $t$. Write the first column $G_0$ as the linear combination of the columns $G_1,\ldots, G_{t}$ which can be written as the following recursive equations 
    \begin{equation}\label{Eq-rec-algo}
g_{i} = \lambda_1 g^{q^u}_{i-1} + \lambda_2 g^{q^{2u} }_{i-2} + \cdots + \lambda_t g^{q^{tu}}_{i-t}, \quad 0\leq i < n.
\end{equation}
\item[(5)] Find the remaining coefficients $g_i$ using the recursive equation \eqref{Eq-rec-algo}. 
\item[(6)] Compute $\tilde{f}=(\beta_0,\ldots,\beta_{n-1})-(g_0,\ldots,g_{n-1})$.
\item[(7)] Compute the codeword $c=\tilde{f}\cdot \mathcal{M^T}^{-1}$.
\end{itemize}


The lines (1) and (7) in the above procedure need $\mathcal{O}(n^3)$ operations over $\mathbb{F}_{q^{un}}$ which can be optimized if one applies the ideas in \cite{PUCHINGER2018194}. The line (2) needs linear complexity while the line (3) dominates the complexity of the whole process. BM algorithm has complexity in the order of $\mathcal{O}(n^2)$ operations over $\mathbb{F}_{q^{un}}$. The complexity of the the remaining steps can be neglected.
\subsection{Examples}
\begin{example}[Symmetric $d$-Codes]\label{Ex:sym}
Let $C$ be an optimal symmetric $d$-code with minimum distance  $d=5$ and length $n=7$ defined over $\mathbb{F}_{2^7}$. We consider a normal basis of $\mathbb{F}_{2^7}$ over $\mathbb{F}_2$ with normal element $w=z^{95}$ as the evaluation points. Here $z$ is a primitive element in $\mathbb{F}_{2^7}^*$

\begin{enumerate}
    \item[]\textbf{Encoding:} Suppose Alice wants to transfer the message $f=(f_0,f_1)=(z^{7}, z^{13})$ to Bob via a noisy channel. The code's evaluation polynomial would have the coefficient vector $\tilde{f}=(f_0,f_1,0,0,0,0,f_1^{q^{6}})=(z^{7}, z^{13}, 0, 0, 0, 0, z^{70})$ which gives the codeword \begin{align*}
  c=\tilde{f}(\mathcal{M}^T)=(z^{108}, z^{36}, z^{11}, z^{12}, z^{57}, z^{24}, z)=\begin{pmatrix}
1 &0& 1& 1& 1& 1& 0\\
0& 0 &1 &0 &0 &1 &0\\
1& 1& 0& 0& 1& 0& 1\\
1& 0& 0& 0& 0& 0& 1\\
1& 0& 1& 0& 1& 0& 1\\
1 &1& 0& 0& 0& 0&0\\
0& 0& 1& 1& 1& 0& 1\\
\end{pmatrix},
     \end{align*}
     in symmetric form.
     \item[]\textbf{Channel's transmission:} We assume that the noisy channel adds an error vector $e=( z^{63}, z^{126}, z^{126}, z^{63}, z^{126}, z^{126}, z^{126})$ with rank $t=2$ to the codeword $c$ and Bob receives the word $$r=c+e=(z^{4}, z^{45}, z^{124}, z^{52}, z^{37}, z^{104}, z^{13}).$$
    \item[]\textbf{Decoding:} Now Bob received $r$ and he wants to recover the message $f$. He first computes $\beta=r\cdot (\mathcal{M}^T)^{-1}=(z^{17}, z^{51}, z^{98}, z^{124}, z^{100}, z^{83}, z^{86} )$ and directly gets the coefficients $(g_2,g_3,g_4,g_5)=(\beta_2,\beta_3,\beta_4,\beta_5)$ where $g(x)=\sum\limits_{i=0}^{6}g_ix^{2^i}$ is the error interpolation polynomial and $\tilde{g}=(g_0,\ldots,g_{6})$. Then he submits $(g_2,g_3,g_4,g_5)$ in the BM algorithm and obtains the unique connection vector $(\lambda_1,\lambda_2)=(z^{25},z^{126})$. Now he uses both $(g_2,g_3,g_4,g_5)$ and $(\lambda_1,\lambda_2)$  as inputs for modified version of LFSR described in \cite{Sidorenk} and get the vector $$a=(g_2,g_3,g_4,g_5,g_6,g_0,g_1)=(z^{98}, z^{124}, z^{100}, z^{83}, z^{55}, z^{115}, z^{71}).$$
    Now he can rearrange the components of $a$ and gets $\tilde{g}=(z^{115}, z^{71},z^{98}, z^{124}, z^{100}, z^{83}, z^{55})$. Since he knows $\beta$ and $\tilde{g}$, he is able to compute $\tilde{f}=\beta-\tilde{g}=(z^{7}, z^{13}, 0, 0, 0, 0, z^{70})$ and finally $f=(\tilde{f}_0,\tilde{f}_1)=(z^7,z^{13})$. 
\end{enumerate}
\end{example}
\begin{example}[Alternating $d$-Codes]\label{Ex:alt}

Suppose $D\in \mathbb{F}_{2^{9}}$ be an alternating $d$-code with length $n=9$ and minimum distance $d=6$. Let $w=z^{347}$ be the normal element for our normal basis which is used as the interpolation points. For the received word $r=(z^{293}, z^{389}, z^{430}, z^{227}, z^{481}, z^{445}, z^{426}, z^{404}, z^{339})$ containing error of $t=\lfloor(d-1)/2\rfloor=2$ rank, we can compute $\beta, (\lambda_1,\lambda_2),a, \tilde{g}, \tilde{f}, c $ and $f$ similar to Example \ref{Ex:sym} as follows: 
\begin{itemize}
    \item $\beta=(z^{486}, z^{233}, z^{334}, z^{155}, z^{167}, z^{226}, z^{483}, z^{231}, z^{88})$,
    \item $(\beta_0,\beta_1,\beta_2,\beta_7,\beta_8)=(g_0,g_1,g_2,g_7,g_8)$. 
    \item BM algorithm input  $(\beta_7,\beta_8,\beta_0,\beta_1,\beta_2)$ gives $(\lambda_1,\lambda_2)=(z^{154}, z^{262})$,
    \item modified LFSR input $(\beta_7,\beta_8,\beta_0,\beta_1,\beta_2)$ and $(\lambda_1,\lambda_2)$ gives
    \begin{align*}
  a&=(\beta_7,\beta_8,\beta_0,\beta_1,\beta_2, g_3,g_4,g_5,g_6)\\&=(z^{231}, z^{88}, z^{486}, z^{233}, z^{334}, z^{505}, z^{113}, z^{265}, z^{425}),
  \end{align*}
  \item $\tilde{g}=(z^{486}, z^{233}, z^{334}, z^{505}, z^{113}, z^{265}, z^{425},z^{231}, z^{88})$,
  \item $\tilde{f}=\beta - \tilde{g}=(0, 0, 0, z^{77}, z^{397}, z^{440}, z^{329}, 0, 0)$,
  \item $c=\tilde{f}\cdot (\mathcal{M}^T)=(z^{244}, z^{412}, z^{364}, z^{400}, z^{368}, z^{161}, z^{122}, z^{59}, z^{122})$,
  \item $f=(f_0,f_1)=(\tilde{f}_{3},\tilde{f}_4)=(z^{77}, z^{397})$.
\end{itemize}
\end{example}
\begin{example}[Hermitian $d$-Codes]\label{Ex:Her2}
Suppose $\mathcal{C}\in \mathbb{F}_{2^{14}}^7$ be an optimal Hermitian $d$-code with length $n=7$, minimum distance $d=5$ and $\eta=z$ . We use  the normal basis $W$ of $\mathbb{F}_{2^{14}}$ over $\mathbb{F}_{2^{2}}$ with normal element $w=z^{8591}$ as the evaluation points, where $z$ is the primitive element in $\mathbb{F}_{2^{10}}^*$. Let $r=(z^{3672}, z^{2957}, z^{1343}, z^{3039}, z^{10923}, z^{9913}, z^{1618})$ be a received word with error of $t=(d-1)/2=2$ rank. Then $\beta=r\cdot (\mathcal{M}^T)^{-1}= (z^{5036}, z^{5234}, z^{203}, z^{840}, z^{2939}, z^{13080}, z^{15830})$. Let $g(x)=\sum\limits_{i=0}^{6}g_ix^{2^{2i}}$ be the error interpolation polynomial. Due to the expected form of $\tilde{f}$ in optimal Hermitian $d$-codes we have  $(\beta_0,\beta_1,\beta_2,\beta_6)=(g_0,g_1,g_2,g_6)$. Now we submit $(\beta_6,\beta_0,\beta_1,\beta_2)$ in the BM algorithm and get the output $(\lambda_1,\lambda_2)=(z^{11141},z^{14283})$. using both $(\beta_6,\beta_0,\beta_1,\beta_2)$ and $(\lambda_1,\lambda_2)$ as the input for the modified version of linear feedback shift register explained in \cite{Sidorenk}, we get $$a=(\beta_6,\beta_0,\beta_1,\beta_2,g_3,g_4,g_5)=(z^{15830}, z^{5036}, z^{5234}, z^{203}, z^{12223}, z^{9784}, z^{1048}).$$
So $\tilde{g}=( z^{5036}, z^{5234}, z^{203}, z^{12223}, z^{9784}, z^{1048},z^{15830})$ and the code's evaluation polynomial has the coefficient vector $\tilde{f}=(0,0,0,z^{4446},z^{11481},z^{15498},0)$. Then the codeword is 

\begin{align*}
c=&\tilde{f}\cdot \mathcal{M}^T=(z^{781}, z^{1313}, z^{4481}, z^{5130}, z^{1671}, z^{9656}, z^{1567})\\
=& \begin{pmatrix}
1 &0& 0& 1& y& y^2& 1\\
0& 1 &0 &y &y &y &1\\
0& 0& 1& 0& y^2& 0& y^2\\
1& y^2& 0& 0& y^2& y^2& 0\\
y^2& y^2& y& y& 1& y& y^2\\
y &y^2& 0& y& y^2& 1&1\\
1& 1& y& 0& y& 1& 1\\
\end{pmatrix},
\end{align*}

where $y$ is the primitive element in $\mathbb{F}_{2^2}^*$ and the message is $f=(f_0,\ldots,f_{k-1})=(l^{89},l^{97},l^{32})$ where $l$ is the primitive element in $\mathbb{F}_{2^7}^*$.
\end{example}

\section{Conclusion}\label{sec:concl}
This work proposes the first encoding and decoding methods for three restricted families of rank metric codes including optimal symmetric, optimal alternating and optimal Hermitian rank metric codes. We showed that the evaluation encoding is a right choice for the aforementioned families  and the proposed encoding methods are easily reversible and efficient. We also introduce  three interpolation-based decoding algorithms  that are based on the  properties of Dickson matrix associated with linearized polynomials. In the decoding process we reduced the rank decoding problem to the problem of solving a system of linear equations which can be solved by Gaussian elimination method or Berlekamp-Massey algorithm in polynomial time. 

\bibliographystyle{abbrv}
\bibliography{RankMetricCodes}


\end{document}